\documentclass[apl,aip,11pt,superscriptaddress,citeautoscript,reprint, notitlepage]{revtex4-1}

\usepackage{amsmath}    
\usepackage{graphicx}   
\usepackage{verbatim}   
\usepackage{pdftexcmds}
\usepackage[version=3]{mhchem} 
\usepackage[T1]{fontenc} 
\usepackage{textgreek}
\usepackage{longtable}  
\usepackage{xcolor}      
\usepackage{subfigure}  
\usepackage{float}      
\usepackage{hyperref}   
\usepackage[latin1]{inputenc}
\usepackage{array} 
\usepackage{tikz}
\usetikzlibrary{shapes,arrows}
\definecolor{cellc}{RGB}{2,159,228}
\definecolor{linc}{RGB}{34,64,116}

\usepackage{array}

\begin{document}

\title{Effect of aging and Mn substitution on anisotropy of third generation piezoelectrics}
\author{Rajasekarakumar Vadapoo} 
\affiliation{Extreme Materials Initiative, Geophysical Laboratory, Carnegie Institution of Washington, Washington, DC 20015, USA}
\author{Muhtar Ahart}
\affiliation{Extreme Materials Initiative, Geophysical Laboratory, Carnegie Institution of Washington, Washington, DC 20015, USA}
\author{Margo Staruch}
\affiliation{Naval Research Laboratory, Washington DC 20375, USA}
\author{Michael Guerette}
\affiliation{Extreme Materials Initiative, Geophysical Laboratory, Carnegie Institution of Washington, Washington, DC 20015, USA}
\author{Jun Luo}
\affiliation{TRS Technologies, Inc., 2820 E. College Ave., State College, Pennsylvania 16801, USA}
\author{Peter Finkel}
\affiliation{Naval Research Laboratory, Washington DC 20375, USA}
\author{R. E. Cohen}
\affiliation{Extreme Materials Initiative, Geophysical Laboratory, Carnegie Institution of Washington, Washington, DC 20015, USA}
\affiliation{Department of Earth and Environmental Sciences, Ludwig Maximilians University Munich, Germany}
\date{\today}

\begin{abstract}
We study the aging and Mn doping effect on third generation lead based relaxor single crystals. We measured the polarization (PE) and strain with applied field on two perpendicular orientations of the rhombohedral pseudocubic [001] poled crystal. We do not find significant ageing or a double hysteresis loop in these samples, either due to the absense of defect dipoles or to the orientation of the samples relative to the polarization direction.

\end{abstract}

\maketitle
Aging in ferroelectrics is found to induce large recoverable strain properties, which could be explored for its effective applications in the field of ferroelectrics and piezoelectric sensors.\cite{Jonker1972, Ren2004} Barium titanate single crystal exhibits a very large recoverable nonlinear strain with aging , which is an order of magnitude higher than the highly strained Pb(Zn$_{1/3}$Nb$_{2/3}$)O$_{3}$ - PbTiO$_{3}$ (PZN-PT) PZN-PT single crystals.\cite{Ren2004} Transition metal dopants are also found to increase its electromechanical properties.\cite{Delfin2011,Wu2008,Nossa2015,hentati2010,park2001} Defect dipoles and aging greatly influence the electromechanical coupling, leading to large changes in cohesive field, strain and piezoelectric properties.\cite{Genenko2015,Guo2015,Li2013,Li2015,Nossa2015,Murakami2015,Zhang2004a,Zhang2005,Zhang2006a}

Relaxor based single crystals like Pb(In$_{1/2}$Nb$_{1/2}$)-PbTiO$_{3}$ (PIN-PT), Pb(In$_{1/2}$Nb$_{1/2}$)O$_{3}$-Pb(Mg$_{1/3}$Nb$_{2/3}$)O$_{3}$-PbTiO$_{3}$ (PIN-PMN-PT) and their acceptor doping have been studied extensively due to their excellent electromechanical properties.\cite{Park1997,Li2011,Zhang2011, Damjanovic2003,Huo2012a,Liu2010,Liu2009,Zhang2012g,Luo2016} Domain engineering in relaxor-PT  crystals show strong anisotripic behaviors when poled along different crystallographic directions. Understanding the underlying mechanism on the influence of defect dipoles due to aging on the undoped and acceptor doped systems of these relaxor materials is crucial. Also, their crystallographic orientation dependence  on electromechanical properties have serious implications for the device applications.

Here we study the effect of  crystals anisotropy coupled with aging on defect-dipole assisted reversible switching of 3rd generation piezoelectrics, Pb(In$_{1/2}$Nb$_{1/2}$)O$_{3}$-Pb(Mg$_{1/3}$ Nb$_{2/3}$)O$_{3}$-PbTiO$_{3}$ (PIN-PMN-PT) and Mn doped PIN-PMN-PT (Mn:PIN-PMN-PT) single crystals.

\begin{figure}
\includegraphics[scale=0.5]{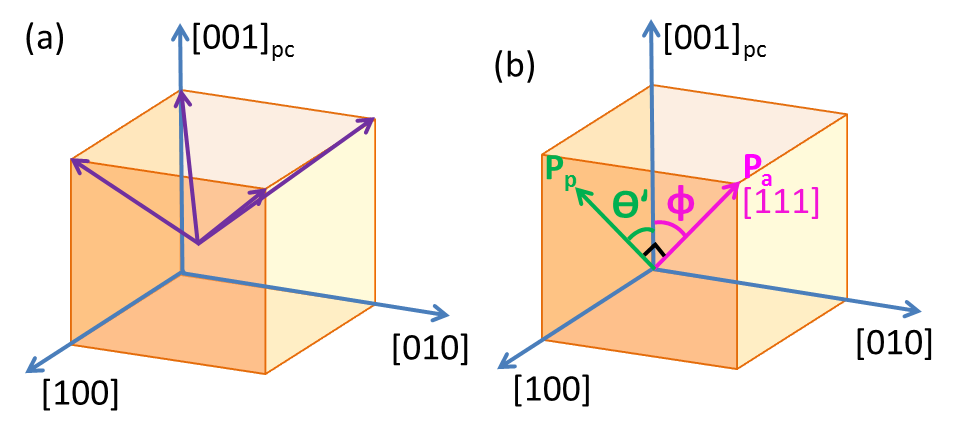} 
\caption{\label{fig:structure} Rhombohedral PIN-PMN-PT and Mn:PIN-PMN-PT crystals within pseudocubic (pc) coordinates (a) [001]$_{pc}$ poled crystal with the possible four equivalent <111> polrarization directions. (b) shows the angular dependence of the average dipole direction to the pseudocubic [001]$_{pc}$ -direction of the crystal. Where, '$\phi$' is the average dipole direction/ average defect dipole direction to the [001]$_{pc}$ direction of the crystal, '$\theta ^{'}$' is the angle between [001]$_{pc}$ and the direction perpendicular to the average dipole direction predicted by our simple model +/- angle of orietnation shift of the crystal calculated from EBSD measurement. }
\end{figure}

\begin{figure}
\includegraphics[scale=0.8]{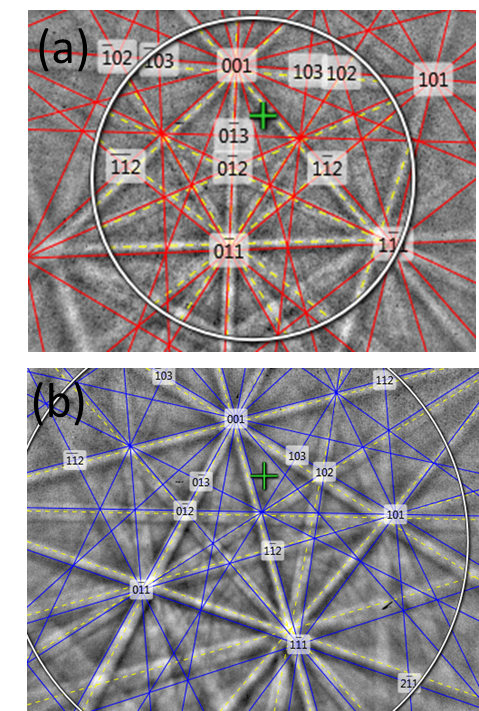} 
\caption{\label{fig:EBSD} Electron back scattered diffraction (EBSD) pattern of [001] pseudo cubic oriented ([001]$_{pc}$) single crystals for (a) PIN-PMN-PT and (b) Mn:PIN-PMN-PT . Yellow lines show the kikutchi  lines unique pattern associate with the crystal diffraction. The red line pattern is the overlay of the crystallographic indexing with the atomic structure of the crystal within the pseudocubic crystalline space group of Pm-3m. The circle line pattern is the area analyzed for the indexing.}
\end{figure}

We examine the rhombohedral [001]$_{pc}$ poled single crystal of PIN-PMN-PT and Mn:PIN-PMN-PT for our studies which were grown by modified Bridgeman method.\cite{Liu2010,Huo2012a} Where, the subscript 'pc' means the pseudocubic coordinates. The [001]$_{pc}$ poled crystal have the four possible equivalent polarization direction along <111> (figure. \ref{fig:structure}. (a)).  The anisotropic and aging effects on polarization and strain with applied electric field on those samples were measured in two perpendicular orientations ([001]$_{pc}$ $\&$ the direction  $\perp _{r}$ to it) of the crystals. We determined the orientation of the crystals by analyzing the kikuchi patterns from electron back scattered diffraction (EBSD) of the PIN-PMN-PT and Mn:PIN-PMN-PT (Figure\ref{fig:EBSD} a) \& b) samples. To record and analyze the EBSD patterns we used the Oxford Nordlys Nano EBSD camera with Aztec pattern recognition software in conjunction with the Jeol JSM-6500F feld emission scanning electron microscope.
The samples were tilted with surface normal being 70$^{\circ}$ to the incident electron beam with the
detector angle at 90$^{\circ}$.  The crystal orientation is determined in terms of euler angles representation within the Bunge notation.\cite{Britton2016}. The crystal structure of these crystals belong to the rhombohedral, R3m space group \ref{Finkel2015,Sun2014} and the EBSD patterns were analyzed using cubic coordinates of the Pm-3m space group with lattice parameter a = 4.0416 \AA. We found that the pseudocubic crystal orientation with the corresponding euler angles of  5.7$^{\circ}$ and  6.6$^{\circ}$ for [001]$_{pc}$ orientated (Z-dir) PIN-PMN-PT and  Mn:PIN-PMN-PT single crystals respectively from the EBSD patterns (figure. \ref{fig:EBSD}).

\begin{figure*}[t]
\includegraphics[scale=0.8]{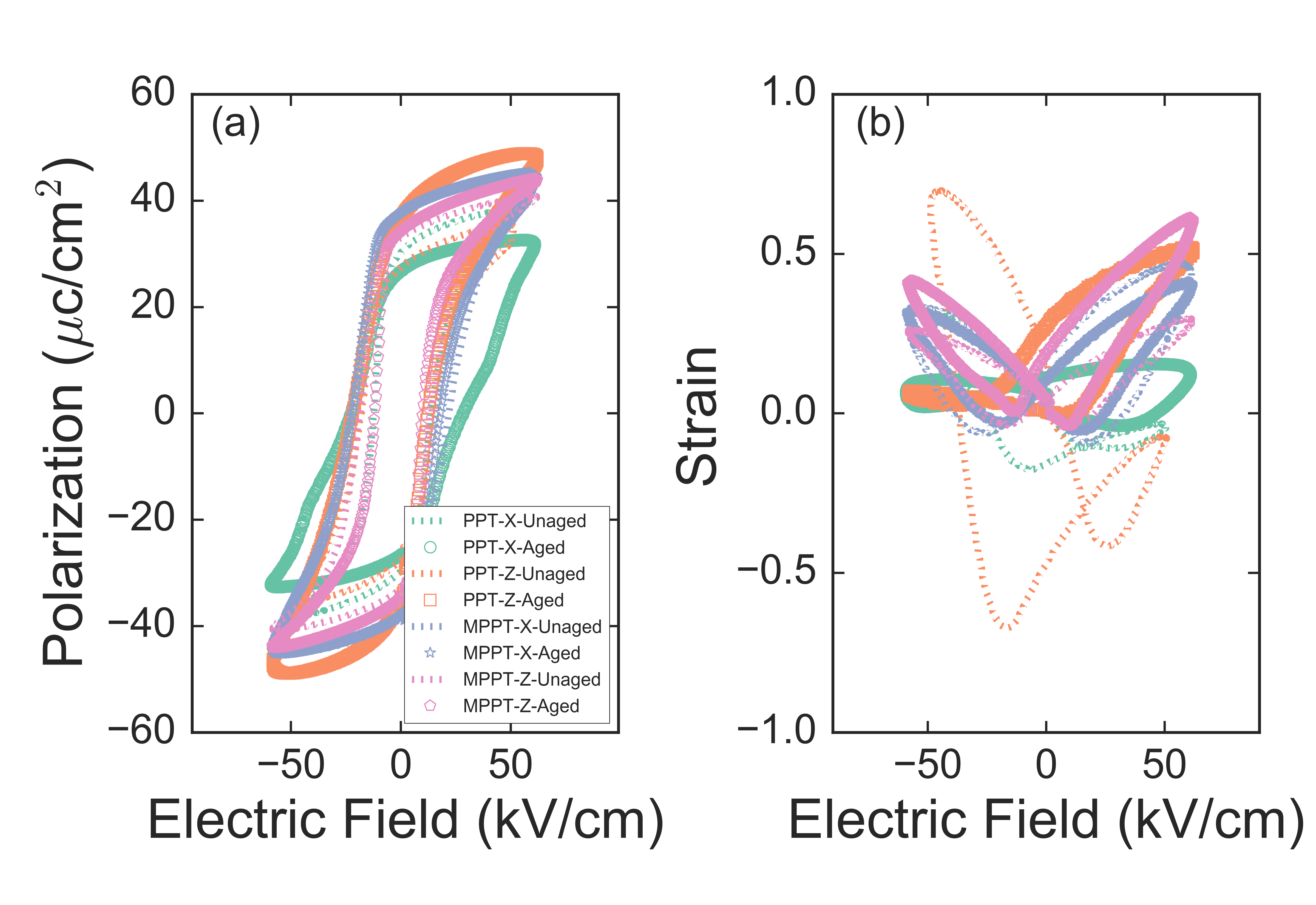} 
\caption{\label{fig:PE-SE} (a) Polarization and (b) Strain of unaged and aged sample of PIN-PMN-PT and Mn:PIN-PMN-PT under applied electric field along two perpendicular orientations ([001]$_{pc}$ orientation (Z-Ori.) \& $\bot_{r}$  to [001]$_{pc}$ orientation (X-Ori.)).}
\end{figure*}

The rhombohedral [001]$_{c}$ poled PIN-PMN-PT and  Mn:PIN-PMN-PT crystals were cut in two perpendicular directions ( [001]$_{pc}$ orientation (Z-Ori.) \& $\bot_{r}$  to [001]$_{pc}$ orientation (X-Ori.)), and reduced to the thickness of $\sim$ 100 microns to perform ferroelectric and strain measurements. Aging on these different oriented crystals were carried out at 85$^{\circ}$ C for 5 days. We measured the effect of polarization and strain with applied electric field along these two mutually perpendicular orientations. We correlate the influence of aging and Mn substitution on  polarization and strain along these two orthogonal orientations (Figure. \ref{fig:PE-SE}).

 The orientation of the crystal and aging have a strong anisotropy in remanent polarization ($P_{r}$) and cohesive field ($E_{c}$) (figure \ref{fig:PE-SE}.(a)). Also, we observed a strong offset of P-E loop with electric field which is a signature of internal bias field similar to the acceptor doped classic ferroelectics BaTiO$_{3}$ and PZT. In PIN-PMN-PT crystal aging reduced the saturation polarization along X-Ori. by 7.03 $\mu C/cm^{2}$ (from 39.67  to 32.64 $\mu C/cm^{2}$) and the cohesive field  significantly increases by 7.5 kV/cm (from $\sim$ 15  to $\sim$ 22.5 kV/cm) and show highly asymmetric behavior with pinching effect (green open circle symbol). Also, in Z-Ori., the polarization increases by 20.58 $\mu C/cm^{2}$ by aging (from 35.55 to 56.13 $\mu C/cm^{2}$) and the cohesive field increases $\sim$ 1 kV/cm (from $\sim$ 15.4  to $\sim$ 16.4 kV/cm). On the other hand, the Mn substitution on PIN-PMN-PT  yielded no change of polarization upon aging in X-Ori.  (45.17 to 45.15  $\mu C/cm^{2}$) with a small reduction of cohesive field of 0.5 kV/cm (from 22.7 to 22.2 kV/cm) with a little as symmetric nature of the hysteresis. In Z-Ori. aging increases the saturation polarization by 3.18 $\mu C/cm^{2}$ (from 40.89 to 44.07$\mu C/cm^{2}$) and decreases the cohesive field by 4 kV/cm (from $\sim$ 16 to $\sim$ 12 kV/cm). The cohesive fields observed on these samples were $\sim$ twice higher than the [001]$_{pc}$ oriented single crystal samples of 1mm thick and higher surface area measured with the same setup, which could be attributed to the surface effect of the crystals and or to the size effect of these 100 micron thickness samples due to the parasitic effect.\cite{Bouregba2003} Overall, aging and Mn substitution was found to increase the remanent polarization in Z-Ori. But in X-Ori. aging  decreased the remanent polarization for PIN-PMN-PT and had no significant change in Mn:PIN-PMN-PT. Also, the X-Ori. PIN-PMN-PT shows the pinching effect of the polarization loop for aging as seen in literature, which we haven't observed in Mn doped samples.

 In PIN-PMN-PT aging significantly reduced the maximum strain (Figure \ref{fig:PE-SE}. (b ) by 0.25 (from 0.40 at -48.38 kV/cm to 0.15 at 47.65 kV/cm), and become less unsymmetric for the X-Ori. Similarly, for Z-Ori. the strain was found to decrease and become more symmetric with aging. Also, in case of Mn substitution the maximum strain decreases by 0.06 (from 0.47 at 57.1 kV/cm to 0.4116 at 60.65 kV/cm) for the X-Ori. with improved symmetric nature of the butterfly loop by aging. On the contrary, in Z-Ori. aging increased the nonsymmetric nature with maximum strain  by 0.31 (0.61 from 0.30 at 60.82 kV/cm). So,it's evident that Mn substitution and aging increased the strain upto a factor of two in Z-Ori.


\bibliography{Aging_Ref1}

\begin{thebibliography}{26}%
\makeatletter
\providecommand \@ifxundefined [1]{%
 \@ifx{#1\undefined}
}%
\providecommand \@ifnum [1]{%
 \ifnum #1\expandafter \@firstoftwo
 \else \expandafter \@secondoftwo
 \fi
}%
\providecommand \@ifx [1]{%
 \ifx #1\expandafter \@firstoftwo
 \else \expandafter \@secondoftwo
 \fi
}%
\providecommand \natexlab [1]{#1}%
\providecommand \enquote  [1]{``#1''}%
\providecommand \bibnamefont  [1]{#1}%
\providecommand \bibfnamefont [1]{#1}%
\providecommand \citenamefont [1]{#1}%
\providecommand \href@noop [0]{\@secondoftwo}%
\providecommand \href [0]{\begingroup \@sanitize@url \@href}%
\providecommand \@href[1]{\@@startlink{#1}\@@href}%
\providecommand \@@href[1]{\endgroup#1\@@endlink}%
\providecommand \@sanitize@url [0]{\catcode `\\12\catcode `\$12\catcode
  `\&12\catcode `\#12\catcode `\^12\catcode `\_12\catcode `\%12\relax}%
\providecommand \@@startlink[1]{}%
\providecommand \@@endlink[0]{}%
\providecommand \url  [0]{\begingroup\@sanitize@url \@url }%
\providecommand \@url [1]{\endgroup\@href {#1}{\urlprefix }}%
\providecommand \urlprefix  [0]{URL }%
\providecommand \Eprint [0]{\href }%
\providecommand \doibase [0]{http://dx.doi.org/}%
\providecommand \selectlanguage [0]{\@gobble}%
\providecommand \bibinfo  [0]{\@secondoftwo}%
\providecommand \bibfield  [0]{\@secondoftwo}%
\providecommand \translation [1]{[#1]}%
\providecommand \BibitemOpen [0]{}%
\providecommand \bibitemStop [0]{}%
\providecommand \bibitemNoStop [0]{.\EOS\space}%
\providecommand \EOS [0]{\spacefactor3000\relax}%
\providecommand \BibitemShut  [1]{\csname bibitem#1\endcsname}%
\let\auto@bib@innerbib\@empty
\bibitem [{\citenamefont {JONKER}(1972)}]{Jonker1972}%
  \BibitemOpen
  \bibfield  {author} {\bibinfo {author} {\bibfnamefont {G.~H.}\ \bibnamefont
  {JONKER}},\ }\href {\doibase 10.1111/j.1151-2916.1972.tb13404.x} {\bibfield
  {journal} {\bibinfo  {journal} {Journal of the American Ceramic Society}\
  }\textbf {\bibinfo {volume} {55}},\ \bibinfo {pages} {57} (\bibinfo {year}
  {1972})}\BibitemShut {NoStop}%
\bibitem [{\citenamefont {Ren}(2004)}]{Ren2004}%
  \BibitemOpen
  \bibfield  {author} {\bibinfo {author} {\bibfnamefont {X.}~\bibnamefont
  {Ren}},\ }\href {\doibase 10.1038/nmat1051} {\bibfield  {journal} {\bibinfo
  {journal} {Nature Materials}\ }\textbf {\bibinfo {volume} {3}},\ \bibinfo
  {pages} {91} (\bibinfo {year} {2004})}\BibitemShut {NoStop}%
\bibitem [{\citenamefont {Perez-Delfin}\ \emph {et~al.}(2011)\citenamefont
  {Perez-Delfin}, \citenamefont {Garcia}, \citenamefont {Ochoa}, \citenamefont
  {Perez}, \citenamefont {Guerrero},\ and\ \citenamefont {Eiras}}]{Delfin2011}%
  \BibitemOpen
  \bibfield  {author} {\bibinfo {author} {\bibfnamefont {E.}~\bibnamefont
  {Perez-Delfin}}, \bibinfo {author} {\bibfnamefont {J.~E.}\ \bibnamefont
  {Garcia}}, \bibinfo {author} {\bibfnamefont {D.~A.}\ \bibnamefont {Ochoa}},
  \bibinfo {author} {\bibfnamefont {R.}~\bibnamefont {Perez}}, \bibinfo
  {author} {\bibfnamefont {F.}~\bibnamefont {Guerrero}}, \ and\ \bibinfo
  {author} {\bibfnamefont {J.~A.}\ \bibnamefont {Eiras}},\ }\href {\doibase
  10.1063/1.3622338} {\bibfield  {journal} {\bibinfo  {journal} {Journal of
  Applied Physics}\ }\textbf {\bibinfo {volume} {110}},\ \bibinfo {pages}
  {034106} (\bibinfo {year} {2011})}\BibitemShut {NoStop}%
\bibitem [{\citenamefont {Wu}\ \emph {et~al.}(2008)\citenamefont {Wu},
  \citenamefont {Wang}, \citenamefont {Chen},\ and\ \citenamefont
  {Wang}}]{Wu2008}%
  \BibitemOpen
  \bibfield  {author} {\bibinfo {author} {\bibfnamefont {S.}~\bibnamefont
  {Wu}}, \bibinfo {author} {\bibfnamefont {S.}~\bibnamefont {Wang}}, \bibinfo
  {author} {\bibfnamefont {L.}~\bibnamefont {Chen}}, \ and\ \bibinfo {author}
  {\bibfnamefont {X.}~\bibnamefont {Wang}},\ }\href {\doibase
  10.1007/s10854-007-9371-9} {\bibfield  {journal} {\bibinfo  {journal}
  {Journal of Materials Science: Materials in Electronics}\ }\textbf {\bibinfo
  {volume} {19}},\ \bibinfo {pages} {505} (\bibinfo {year} {2008})}\BibitemShut
  {NoStop}%
\bibitem [{\citenamefont {Nossa}, \citenamefont {Naumov},\ and\ \citenamefont
  {Cohen}(2015)}]{Nossa2015}%
  \BibitemOpen
  \bibfield  {author} {\bibinfo {author} {\bibfnamefont {J.~F.}\ \bibnamefont
  {Nossa}}, \bibinfo {author} {\bibfnamefont {I.~I.}\ \bibnamefont {Naumov}}, \
  and\ \bibinfo {author} {\bibfnamefont {R.~E.}\ \bibnamefont {Cohen}},\ }\href
  {\doibase 10.1103/PhysRevB.91.214105} {\bibfield  {journal} {\bibinfo
  {journal} {Physical Review B}\ }\textbf {\bibinfo {volume} {91}},\ \bibinfo
  {pages} {214105} (\bibinfo {year} {2015})}\BibitemShut {NoStop}%
\bibitem [{\citenamefont {Hentati}\ \emph {et~al.}(2010)\citenamefont
  {Hentati}, \citenamefont {Guennou}, \citenamefont {Dammak}, \citenamefont
  {Khemakhem},\ and\ \citenamefont {Thi}}]{hentati2010}%
  \BibitemOpen
  \bibfield  {author} {\bibinfo {author} {\bibfnamefont {M.~A.}\ \bibnamefont
  {Hentati}}, \bibinfo {author} {\bibfnamefont {M.}~\bibnamefont {Guennou}},
  \bibinfo {author} {\bibfnamefont {H.}~\bibnamefont {Dammak}}, \bibinfo
  {author} {\bibfnamefont {H.}~\bibnamefont {Khemakhem}}, \ and\ \bibinfo
  {author} {\bibfnamefont {M.~P.}\ \bibnamefont {Thi}},\ }\href {\doibase
  10.1063/1.3331817} {\bibfield  {journal} {\bibinfo  {journal} {Journal of
  Applied Physics}\ }\textbf {\bibinfo {volume} {107}},\ \bibinfo {pages}
  {064108} (\bibinfo {year} {2010})}\BibitemShut {NoStop}%
\bibitem [{\citenamefont {Park}\ \emph {et~al.}(2001)\citenamefont {Park},
  \citenamefont {Park}, \citenamefont {Park}, \citenamefont {Kim},\ and\
  \citenamefont {Kim}}]{park2001}%
  \BibitemOpen
  \bibfield  {author} {\bibinfo {author} {\bibfnamefont {J.-H.}\ \bibnamefont
  {Park}}, \bibinfo {author} {\bibfnamefont {J.}~\bibnamefont {Park}}, \bibinfo
  {author} {\bibfnamefont {J.-g.}\ \bibnamefont {Park}}, \bibinfo {author}
  {\bibfnamefont {B.-K.}\ \bibnamefont {Kim}}, \ and\ \bibinfo {author}
  {\bibfnamefont {Y.}~\bibnamefont {Kim}},\ }\href {\doibase
  10.1016/S0955-2219(01)00023-1} {\bibfield  {journal} {\bibinfo  {journal}
  {Journal of the European Ceramic Society}\ }\textbf {\bibinfo {volume}
  {21}},\ \bibinfo {pages} {1383} (\bibinfo {year} {2001})}\BibitemShut
  {NoStop}%
\bibitem [{\citenamefont {Genenko}\ \emph {et~al.}(2015)\citenamefont
  {Genenko}, \citenamefont {Glaum}, \citenamefont {Hoffmann},\ and\
  \citenamefont {Albe}}]{Genenko2015}%
  \BibitemOpen
  \bibfield  {author} {\bibinfo {author} {\bibfnamefont {Y.~A.}\ \bibnamefont
  {Genenko}}, \bibinfo {author} {\bibfnamefont {J.}~\bibnamefont {Glaum}},
  \bibinfo {author} {\bibfnamefont {M.~J.}\ \bibnamefont {Hoffmann}}, \ and\
  \bibinfo {author} {\bibfnamefont {K.}~\bibnamefont {Albe}},\ }\href {\doibase
  10.1016/j.mseb.2014.10.003} {\bibfield  {journal} {\bibinfo  {journal}
  {Materials Science and Engineering: B}\ }\textbf {\bibinfo {volume} {192}},\
  \bibinfo {pages} {52} (\bibinfo {year} {2015})}\BibitemShut {NoStop}%
\bibitem [{\citenamefont {Guo}\ \emph {et~al.}(2015)\citenamefont {Guo},
  \citenamefont {Liu}, \citenamefont {Guo}, \citenamefont {Wei}, \citenamefont
  {Guo},\ and\ \citenamefont {Zhang}}]{Guo2015}%
  \BibitemOpen
  \bibfield  {author} {\bibinfo {author} {\bibfnamefont {Y.~Y.}\ \bibnamefont
  {Guo}}, \bibinfo {author} {\bibfnamefont {J.-M.}\ \bibnamefont {Liu}},
  \bibinfo {author} {\bibfnamefont {Y.~F.}\ \bibnamefont {Guo}}, \bibinfo
  {author} {\bibfnamefont {T.}~\bibnamefont {Wei}}, \bibinfo {author}
  {\bibfnamefont {Y.~J.}\ \bibnamefont {Guo}}, \ and\ \bibinfo {author}
  {\bibfnamefont {N.}~\bibnamefont {Zhang}},\ }\href {\doibase
  10.1063/1.4930259} {\bibfield  {journal} {\bibinfo  {journal} {AIP Advances}\
  }\textbf {\bibinfo {volume} {5}},\ \bibinfo {pages} {097107} (\bibinfo {year}
  {2015})}\BibitemShut {NoStop}%
\bibitem [{\citenamefont {Li}\ and\ \citenamefont {Li}(2013)}]{Li2013}%
  \BibitemOpen
  \bibfield  {author} {\bibinfo {author} {\bibfnamefont {Y.~W.}\ \bibnamefont
  {Li}}\ and\ \bibinfo {author} {\bibfnamefont {F.~X.}\ \bibnamefont {Li}},\
  }\href {\doibase 10.1063/1.4802728} {\bibfield  {journal} {\bibinfo
  {journal} {Applied Physics Letters}\ }\textbf {\bibinfo {volume} {102}},\
  \bibinfo {pages} {152905} (\bibinfo {year} {2013})}\BibitemShut {NoStop}%
\bibitem [{\citenamefont {Li}\ and\ \citenamefont {Li}(2015)}]{Li2015}%
  \BibitemOpen
  \bibfield  {author} {\bibinfo {author} {\bibfnamefont {Y.~W.}\ \bibnamefont
  {Li}}\ and\ \bibinfo {author} {\bibfnamefont {F.~X.}\ \bibnamefont {Li}},\
  }\href {\doibase 10.1063/1.4922970} {\bibfield  {journal} {\bibinfo
  {journal} {Journal of Applied Physics}\ }\textbf {\bibinfo {volume} {117}},\
  \bibinfo {pages} {244101} (\bibinfo {year} {2015})}\BibitemShut {NoStop}%
\bibitem [{\citenamefont {Murakami}\ \emph {et~al.}(2015)\citenamefont
  {Murakami}, \citenamefont {Watanabe}, \citenamefont {Suzuki}, \citenamefont
  {Matsuda},\ and\ \citenamefont {Miura}}]{Murakami2015}%
  \BibitemOpen
  \bibfield  {author} {\bibinfo {author} {\bibfnamefont {S.}~\bibnamefont
  {Murakami}}, \bibinfo {author} {\bibfnamefont {T.}~\bibnamefont {Watanabe}},
  \bibinfo {author} {\bibfnamefont {T.}~\bibnamefont {Suzuki}}, \bibinfo
  {author} {\bibfnamefont {T.}~\bibnamefont {Matsuda}}, \ and\ \bibinfo
  {author} {\bibfnamefont {K.}~\bibnamefont {Miura}},\ }\href {\doibase
  10.7567/JJAP.54.10ND05} {\bibfield  {journal} {\bibinfo  {journal} {Japanese
  Journal of Applied Physics}\ }\textbf {\bibinfo {volume} {54}},\ \bibinfo
  {pages} {10ND05} (\bibinfo {year} {2015})}\BibitemShut {NoStop}%
\bibitem [{\citenamefont {Zhang}, \citenamefont {Chen},\ and\ \citenamefont
  {Ren}(2004)}]{Zhang2004a}%
  \BibitemOpen
  \bibfield  {author} {\bibinfo {author} {\bibfnamefont {L.~X.}\ \bibnamefont
  {Zhang}}, \bibinfo {author} {\bibfnamefont {W.}~\bibnamefont {Chen}}, \ and\
  \bibinfo {author} {\bibfnamefont {X.}~\bibnamefont {Ren}},\ }\href {\doibase
  10.1063/1.1829394} {\bibfield  {journal} {\bibinfo  {journal} {Applied
  Physics Letters}\ }\textbf {\bibinfo {volume} {85}},\ \bibinfo {pages} {5658}
  (\bibinfo {year} {2004})}\BibitemShut {NoStop}%
\bibitem [{\citenamefont {Zhang}\ and\ \citenamefont {Ren}(2005)}]{Zhang2005}%
  \BibitemOpen
  \bibfield  {author} {\bibinfo {author} {\bibfnamefont {L.~X.}\ \bibnamefont
  {Zhang}}\ and\ \bibinfo {author} {\bibfnamefont {X.}~\bibnamefont {Ren}},\
  }\href {\doibase 10.1103/PhysRevB.71.174108} {\bibfield  {journal} {\bibinfo
  {journal} {Physical Review B}\ }\textbf {\bibinfo {volume} {71}},\ \bibinfo
  {pages} {174108} (\bibinfo {year} {2005})}\BibitemShut {NoStop}%
\bibitem [{\citenamefont {Zhang}\ and\ \citenamefont {Ren}(2006)}]{Zhang2006a}%
  \BibitemOpen
  \bibfield  {author} {\bibinfo {author} {\bibfnamefont {L.}~\bibnamefont
  {Zhang}}\ and\ \bibinfo {author} {\bibfnamefont {X.}~\bibnamefont {Ren}},\
  }\href {\doibase 10.1103/PhysRevB.73.094121} {\bibfield  {journal} {\bibinfo
  {journal} {Physical Review B}\ }\textbf {\bibinfo {volume} {73}},\ \bibinfo
  {pages} {094121} (\bibinfo {year} {2006})}\BibitemShut {NoStop}%
\bibitem [{\citenamefont {Park}\ and\ \citenamefont {Shrout}(1997)}]{Park1997}%
  \BibitemOpen
  \bibfield  {author} {\bibinfo {author} {\bibfnamefont {S.-E.}\ \bibnamefont
  {Park}}\ and\ \bibinfo {author} {\bibfnamefont {T.~R.}\ \bibnamefont
  {Shrout}},\ }\href {\doibase 10.1063/1.365983} {\bibfield  {journal}
  {\bibinfo  {journal} {Journal of Applied Physics}\ }\textbf {\bibinfo
  {volume} {82}},\ \bibinfo {pages} {1804} (\bibinfo {year}
  {1997})}\BibitemShut {NoStop}%
\bibitem [{\citenamefont {Li}\ \emph {et~al.}(2011)\citenamefont {Li},
  \citenamefont {Zhang}, \citenamefont {Lin}, \citenamefont {Luo},
  \citenamefont {Xu}, \citenamefont {Wei},\ and\ \citenamefont
  {Shrout}}]{Li2011}%
  \BibitemOpen
  \bibfield  {author} {\bibinfo {author} {\bibfnamefont {F.}~\bibnamefont
  {Li}}, \bibinfo {author} {\bibfnamefont {S.}~\bibnamefont {Zhang}}, \bibinfo
  {author} {\bibfnamefont {D.}~\bibnamefont {Lin}}, \bibinfo {author}
  {\bibfnamefont {J.}~\bibnamefont {Luo}}, \bibinfo {author} {\bibfnamefont
  {Z.}~\bibnamefont {Xu}}, \bibinfo {author} {\bibfnamefont {X.}~\bibnamefont
  {Wei}}, \ and\ \bibinfo {author} {\bibfnamefont {T.~R.}\ \bibnamefont
  {Shrout}},\ }\href {\doibase 10.1063/1.3530617} {\bibfield  {journal}
  {\bibinfo  {journal} {Journal of Applied Physics}\ }\textbf {\bibinfo
  {volume} {109}},\ \bibinfo {pages} {014108} (\bibinfo {year}
  {2011})}\BibitemShut {NoStop}%
\bibitem [{\citenamefont {Zhang}\ \emph {et~al.}(2011)\citenamefont {Zhang},
  \citenamefont {Liu}, \citenamefont {Jiang}, \citenamefont {Luo},
  \citenamefont {Cao},\ and\ \citenamefont {Shrout}}]{Zhang2011}%
  \BibitemOpen
  \bibfield  {author} {\bibinfo {author} {\bibfnamefont {S.}~\bibnamefont
  {Zhang}}, \bibinfo {author} {\bibfnamefont {G.}~\bibnamefont {Liu}}, \bibinfo
  {author} {\bibfnamefont {W.}~\bibnamefont {Jiang}}, \bibinfo {author}
  {\bibfnamefont {J.}~\bibnamefont {Luo}}, \bibinfo {author} {\bibfnamefont
  {W.}~\bibnamefont {Cao}}, \ and\ \bibinfo {author} {\bibfnamefont {T.~R.}\
  \bibnamefont {Shrout}},\ }\href {\doibase 10.1063/1.3639316} {\bibfield
  {journal} {\bibinfo  {journal} {Journal of Applied Physics}\ }\textbf
  {\bibinfo {volume} {110}},\ \bibinfo {pages} {064108} (\bibinfo {year}
  {2011})}\BibitemShut {NoStop}%
\bibitem [{\citenamefont {Damjanovic}\ \emph {et~al.}(2003)\citenamefont
  {Damjanovic}, \citenamefont {Budimir}, \citenamefont {Davis},\ and\
  \citenamefont {Setter}}]{Damjanovic2003}%
  \BibitemOpen
  \bibfield  {author} {\bibinfo {author} {\bibfnamefont {D.}~\bibnamefont
  {Damjanovic}}, \bibinfo {author} {\bibfnamefont {M.}~\bibnamefont {Budimir}},
  \bibinfo {author} {\bibfnamefont {M.}~\bibnamefont {Davis}}, \ and\ \bibinfo
  {author} {\bibfnamefont {N.}~\bibnamefont {Setter}},\ }\href {\doibase
  10.1063/1.1592880} {\bibfield  {journal} {\bibinfo  {journal} {Applied
  Physics Letters}\ }\textbf {\bibinfo {volume} {83}},\ \bibinfo {pages} {527}
  (\bibinfo {year} {2003})}\BibitemShut {NoStop}%
\bibitem [{\citenamefont {Huo}\ \emph {et~al.}(2012)\citenamefont {Huo},
  \citenamefont {Zhang}, \citenamefont {Liu}, \citenamefont {Zhang},
  \citenamefont {Luo}, \citenamefont {Sahul}, \citenamefont {Cao},\ and\
  \citenamefont {Shrout}}]{Huo2012a}%
  \BibitemOpen
  \bibfield  {author} {\bibinfo {author} {\bibfnamefont {X.}~\bibnamefont
  {Huo}}, \bibinfo {author} {\bibfnamefont {S.}~\bibnamefont {Zhang}}, \bibinfo
  {author} {\bibfnamefont {G.}~\bibnamefont {Liu}}, \bibinfo {author}
  {\bibfnamefont {R.}~\bibnamefont {Zhang}}, \bibinfo {author} {\bibfnamefont
  {J.}~\bibnamefont {Luo}}, \bibinfo {author} {\bibfnamefont {R.}~\bibnamefont
  {Sahul}}, \bibinfo {author} {\bibfnamefont {W.}~\bibnamefont {Cao}}, \ and\
  \bibinfo {author} {\bibfnamefont {T.~R.}\ \bibnamefont {Shrout}},\ }\href
  {\doibase 10.1063/1.4772617} {\bibfield  {journal} {\bibinfo  {journal}
  {Journal of Applied Physics}\ }\textbf {\bibinfo {volume} {112}},\ \bibinfo
  {pages} {124113} (\bibinfo {year} {2012})}\BibitemShut {NoStop}%
\bibitem [{\citenamefont {Liu}\ \emph {et~al.}(2010)\citenamefont {Liu},
  \citenamefont {Zhang}, \citenamefont {Luo}, \citenamefont {Shrout},\ and\
  \citenamefont {Cao}}]{Liu2010}%
  \BibitemOpen
  \bibfield  {author} {\bibinfo {author} {\bibfnamefont {X.}~\bibnamefont
  {Liu}}, \bibinfo {author} {\bibfnamefont {S.}~\bibnamefont {Zhang}}, \bibinfo
  {author} {\bibfnamefont {J.}~\bibnamefont {Luo}}, \bibinfo {author}
  {\bibfnamefont {T.~R.}\ \bibnamefont {Shrout}}, \ and\ \bibinfo {author}
  {\bibfnamefont {W.}~\bibnamefont {Cao}},\ }\href {\doibase 10.1063/1.3275803}
  {\bibfield  {journal} {\bibinfo  {journal} {Applied Physics Letters}\
  }\textbf {\bibinfo {volume} {96}},\ \bibinfo {pages} {012907} (\bibinfo
  {year} {2010})}\BibitemShut {NoStop}%
\bibitem [{\citenamefont {Liu}\ \emph {et~al.}(2009)\citenamefont {Liu},
  \citenamefont {Zhang}, \citenamefont {Luo}, \citenamefont {Shrout},\ and\
  \citenamefont {Cao}}]{Liu2009}%
  \BibitemOpen
  \bibfield  {author} {\bibinfo {author} {\bibfnamefont {X.}~\bibnamefont
  {Liu}}, \bibinfo {author} {\bibfnamefont {S.}~\bibnamefont {Zhang}}, \bibinfo
  {author} {\bibfnamefont {J.}~\bibnamefont {Luo}}, \bibinfo {author}
  {\bibfnamefont {T.~R.}\ \bibnamefont {Shrout}}, \ and\ \bibinfo {author}
  {\bibfnamefont {W.}~\bibnamefont {Cao}},\ }\href {\doibase 10.1063/1.3243169}
  {\bibfield  {journal} {\bibinfo  {journal} {Journal of Applied Physics}\
  }\textbf {\bibinfo {volume} {106}},\ \bibinfo {pages} {074112} (\bibinfo
  {year} {2009})}\BibitemShut {NoStop}%
\bibitem [{\citenamefont {Zhang}\ and\ \citenamefont {Li}(2012)}]{Zhang2012g}%
  \BibitemOpen
  \bibfield  {author} {\bibinfo {author} {\bibfnamefont {S.}~\bibnamefont
  {Zhang}}\ and\ \bibinfo {author} {\bibfnamefont {F.}~\bibnamefont {Li}},\
  }\href {\doibase 10.1063/1.3679521} {\bibfield  {journal} {\bibinfo
  {journal} {Journal of Applied Physics}\ }\textbf {\bibinfo {volume} {111}},\
  \bibinfo {pages} {031301} (\bibinfo {year} {2012})}\BibitemShut {NoStop}%
\bibitem [{\citenamefont {Luo}\ \emph {et~al.}(2016)\citenamefont {Luo},
  \citenamefont {Zhang}, \citenamefont {Li}, \citenamefont {Yan}, \citenamefont
  {Zhang}, \citenamefont {Ansell}, \citenamefont {Luo},\ and\ \citenamefont
  {Shrout}}]{Luo2016}%
  \BibitemOpen
  \bibfield  {author} {\bibinfo {author} {\bibfnamefont {N.}~\bibnamefont
  {Luo}}, \bibinfo {author} {\bibfnamefont {S.}~\bibnamefont {Zhang}}, \bibinfo
  {author} {\bibfnamefont {Q.}~\bibnamefont {Li}}, \bibinfo {author}
  {\bibfnamefont {Q.}~\bibnamefont {Yan}}, \bibinfo {author} {\bibfnamefont
  {Y.}~\bibnamefont {Zhang}}, \bibinfo {author} {\bibfnamefont
  {T.}~\bibnamefont {Ansell}}, \bibinfo {author} {\bibfnamefont
  {J.}~\bibnamefont {Luo}}, \ and\ \bibinfo {author} {\bibfnamefont {T.~R.}\
  \bibnamefont {Shrout}},\ }\href {\doibase 10.1039/C6TC00875E} {\bibfield
  {journal} {\bibinfo  {journal} {J. Mater. Chem. C}\ }\textbf {\bibinfo
  {volume} {4}},\ \bibinfo {pages} {4568} (\bibinfo {year} {2016})}\BibitemShut
  {NoStop}%
\bibitem [{\citenamefont {Britton}\ \emph {et~al.}(2016)\citenamefont
  {Britton}, \citenamefont {Jiang}, \citenamefont {Guo}, \citenamefont
  {Vilalta-Clemente}, \citenamefont {Wallis}, \citenamefont {Hansen},
  \citenamefont {Winkelmann},\ and\ \citenamefont {Wilkinson}}]{Britton2016}%
  \BibitemOpen
  \bibfield  {author} {\bibinfo {author} {\bibfnamefont {T.}~\bibnamefont
  {Britton}}, \bibinfo {author} {\bibfnamefont {J.}~\bibnamefont {Jiang}},
  \bibinfo {author} {\bibfnamefont {Y.}~\bibnamefont {Guo}}, \bibinfo {author}
  {\bibfnamefont {A.}~\bibnamefont {Vilalta-Clemente}}, \bibinfo {author}
  {\bibfnamefont {D.}~\bibnamefont {Wallis}}, \bibinfo {author} {\bibfnamefont
  {L.}~\bibnamefont {Hansen}}, \bibinfo {author} {\bibfnamefont
  {A.}~\bibnamefont {Winkelmann}}, \ and\ \bibinfo {author} {\bibfnamefont
  {A.}~\bibnamefont {Wilkinson}},\ }\href {\doibase
  10.1016/j.matchar.2016.04.008} {\bibfield  {journal} {\bibinfo  {journal}
  {Materials Characterization}\ }\textbf {\bibinfo {volume} {117}},\ \bibinfo
  {pages} {113} (\bibinfo {year} {2016})}\BibitemShut {NoStop}%
\bibitem [{\citenamefont {Bouregba}\ \emph {et~al.}(2003)\citenamefont
  {Bouregba}, \citenamefont {Vilquin}, \citenamefont {{Le Rhun}}, \citenamefont
  {Poullain},\ and\ \citenamefont {Domenges}}]{Bouregba2003}%
  \BibitemOpen
  \bibfield  {author} {\bibinfo {author} {\bibfnamefont {R.}~\bibnamefont
  {Bouregba}}, \bibinfo {author} {\bibfnamefont {B.}~\bibnamefont {Vilquin}},
  \bibinfo {author} {\bibfnamefont {G.}~\bibnamefont {{Le Rhun}}}, \bibinfo
  {author} {\bibfnamefont {G.}~\bibnamefont {Poullain}}, \ and\ \bibinfo
  {author} {\bibfnamefont {B.}~\bibnamefont {Domenges}},\ }\href {\doibase
  http://dx.doi.org/10.1063/1.1606533} {\bibfield  {journal} {\bibinfo
  {journal} {Review of Scientific Instruments}\ }\textbf {\bibinfo {volume}
  {74}},\ \bibinfo {pages} {4429} (\bibinfo {year} {2003})}\BibitemShut
  {NoStop}%
\end{thebibliography}%

\end{document}